%% file: main.tex
\documentclass[lettersize,journal]{IEEEtran}
\usepackage{amsmath,amsfonts}
\usepackage{algorithmic}
\usepackage{algorithm}
\usepackage{array}
\usepackage{textcomp}
\usepackage{stfloats}
\usepackage{url}
\usepackage{verbatim}
\usepackage{graphicx}
\usepackage{float}  
\usepackage{cite}
\usepackage{booktabs}
\usepackage{multirow}
\usepackage{subfigure}  %
\begin{document}

\title{Joint Multiscale Cross-lingual Speaking Style Transfer with Bidirectional Attention Mechanism for Automatic Dubbing}

\author{
    Jingbei Li,
    Sipan Li,
    Ping Chen,
    Luwen Zhang,
    Yi Meng,
    Zhiyong Wu,~\IEEEmembership{Member, ~IEEE,}\\
    Helen Meng,~\IEEEmembership{Fellow, ~IEEE,}
    Qiao Tian,
    Yuping Wang,
    Yuxuan Wang
\thanks{
Manuscript received 27 January 2023; revised 19 August 2023 and 12 October 2023; accepted 3 November 2023. This work was supported in part by the National Natural Science Foundation of China under Grant 62076144, in part by the Shenzhen Science and Technology Innovation Committee under Grant WDZC20220816140515001, and in part by the AMiner. Shenzhen SciBrain fund and Shenzhen Key Laboratory of next-generation interactive media innovative technology under Grant ZDSYS20210623092001004. The associate editor coordinating the review of this manuscript and approving it for publication was Prof. Joseph Keshet. \textit{(Jingbei Li and Sipan Li equally contributed to this work.)} \textit{(Corresponding author: Zhiyong Wu.)}}%
\thanks{Jingbei Li, Sipan Li, Ping Chen, Luwen Zhang, Yi Meng and Zhiyong Wu are with
the Tsinghua-CUHK Joint Research Center for Media Sciences, Technologies and Systems,
Shenzhen International Graduate School, Tsinghua University, Shenzhen 518055, China
(e-mail:
lijb19@mails.tsinghua.edu.cn;
lisp20@mails.tsinghua.edu.cn;
p-chen21@mails.tsinghua.edu.cn;
zlw20@\\mails.tsinghua.edu.cn;
my20@mails.tsinghua.edu.cn;
zywu@sz.tsinghua.edu.\\cn).}
\thanks{Helen Meng is with
the Department of Systems Engineering and Engineering Management,
The Chinese University of Hong Kong, Hong Kong SAR, China
(e-mail: hmmeng@se.cuhk.edu.hk).}%
\thanks{Qiao Tian, Yuping Wang and Yuxuan Wang are with
ByteDance, Shanghai, China.
(e-mail:
tianqiao.wave@bytedance.com;
wangyuping@bytedance.com;
wangyuxuan.11@bytedance.com).}
\thanks{Digital Object Identifier 10.1109/TASLP.2023.3331813}
}

\markboth{Journal of \LaTeX\ Class Files,~Vol.~14, No.~8, August~2021}%
{Shell \MakeLowercase{\textit{Jingbei Li et al.}}: Joint Multiscale Cross-lingual Speaking Style Transfer with Bidirectional Attention Mechanism for Automatic Dubbing}

\maketitle

\begin{abstract}
Automatic dubbing, which generates a corresponding version of the input speech in another language, can be widely utilized in many real-world scenarios, such as video and game localization. In addition to synthesizing the translated scripts, automatic dubbing further transfers the speaking style in the original language to the dubbed speeches to give audiences the impression that the characters are speaking in their native tongue. However, state-of-the-art automatic dubbing systems only model the transfer on the duration and speaking rate, disregarding the other aspects of speaking style, such as emotion, intonation and emphasis, which are also crucial to fully understand the characters and speech. In this paper, we propose a joint multiscale cross-lingual speaking style transfer framework to simultaneously model the bidirectional speaking style transfer between two languages at both the global scale (i.e., utterance level) and local scale (i.e., word level). The global and local speaking styles in each language are extracted and utilized to predict the global and local speaking styles in the other language with an encoder-decoder framework for each direction and a shared bidirectional attention mechanism for both directions. A multiscale speaking style-enhanced FastSpeech 2 is then utilized to synthesize the desired speech with the predicted global and local speaking styles for each language. The experimental results demonstrate the effectiveness of our proposed framework, which outperforms a baseline with only duration transfer in objective and subjective evaluations.
\end{abstract}

\newpage


\begin{figure*}[t]
	\centering
	\includegraphics[width=.7\linewidth]{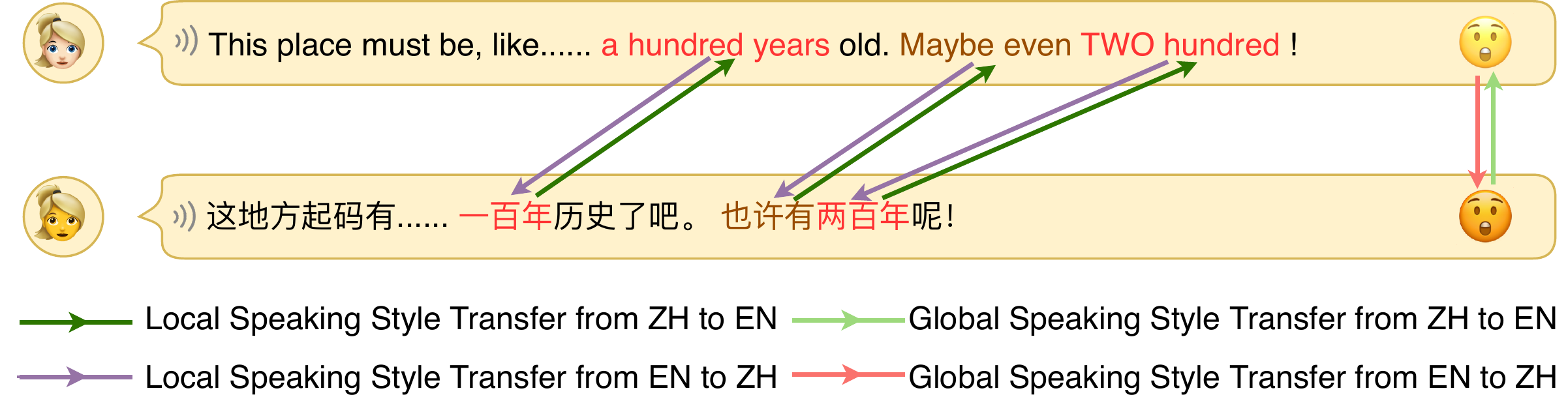}
	\caption{Cross-lingual speaking style transfer between two languages at multiple scales.}
	\label{fig:emoji}
\end{figure*}

\section{Introduction}

\IEEEPARstart{T}{ext}-TO-SPEECH (TTS) synthesis, 
which aims to generate natural speech from given text \cite{taylor2009text},
has been widely researched and utilized in many real-world applications.
With the development of speech processing and neural network technologies,
state-of-the-art TTS systems \cite{wang2017tacotron, shen2018natural, li2019neural, ren2020fastspeech, kim2021conditional} have achieved end-to-end speech synthesis
by merging different components in conventional TTS systems into one trainable framework
to synthesize high-quality speeches while requiring less feature engineering.
Such end-to-end neural network-based TTS models have been successfully implemented in various languages \cite{li2019knowledge},
and widely employed as the backbones for many TTS-related tasks, %
such as emotional speech synthesis \cite{cai2021emotion, lei2022msemotts}, speaking style transfer \cite{wang2018style, li2021towards, li2022towards}, cross-lingual TTS systems \cite{sun2016personalized, tu2019end, chen2019cross} and automatic dubbing \cite{oktem2019prosodic}.

Among these tasks, %
automatic dubbing \cite{oktem2019prosodic}, %
which converts the
speeches in films, televisions or games to
their corresponding versions in a different language rendition,
is a potential research area
and can be utilized in film and game localization to simplify the laborious human dubbing process.
Different from standard TTS synthesis \cite{wang2017tacotron, shen2018natural, li2019neural, ren2020fastspeech, kim2021conditional}
which %
directly synthesize speech from given text,
automatic dubbing needs to further consider the speaking styles in speech
and transfer them to the dubbed speech during synthesis.
In film or game production, actors use various speaking styles to
present the personalities, emotions and intentions of different characters,
including the speaking styles at the global scale (i.e., utterance level) and local scale (i.e., word level).
Global speaking style \cite{wang2018style} is used to control the global emotion and intention for each utterance,
while local speaking style \cite{li_inferring_2022} is used to control the emphasis and intonation for each word to highlight the details in each utterance.
Both kinds of speaking styles are
crucial for speech understanding
and should be transferred in automatic dubbing
to give audiences the impression that the characters are speaking in their native tongue \cite{brannon_dubbing_2023}.

State-of-the-art automatic dubbing systems \cite{9747158} have successfully transferred speaking styles at 
prosodic phrase level.
The studies employ algorithms based on attention mechanisms \cite{oktem2019prosodic} or hidden Markov models (HMMs) \cite{federico2020evaluating, virkar2022prosodic} 
to match the prosodic phrases in the two languages.
Then, several prosodic phrase alignment methods \cite{oktem2019prosodic} are proposed to adjust the duration \cite{9747158}, speaking rate,
or speaking style \cite{swiatkowski_expressive_2023}
for each prosodic phrase \cite{virkar2021improvements} to match the duration of the corresponding prosodic phrase in the speech in the original language.

However, the current automatic dubbing systems
are typically suitable for only a limited range of simple dubbing scenarios and have insufficient modeling on multiscale speaking style transfer.
These approaches commonly require consistency between the number of prosodic phrases in the dubbing language and those in the source language. %
For complicated scenarios, where the number of prosodic phrases is often different, such as dubbing between Chinese and English,
these approaches %
may synthesize improper pauses and word durations. %
Furthermore, these approaches
can only adjust the speaking style of %
each prosodic phrase according to the prosodic phrases in the other language at the same position, disregarding whether they have the same semantic meanings.
Such a design can produce improper word durations and speaking rates when the orders of prosodic phrases are significantly different in the original and dubbed languages.
Moreover, %
the modeling of speaking style transfer at the global scale is also rarely considered in existing approaches. %
The transfer of global speaking style is vital for utterances that are sense-for-sense translation, in which the meanings of each word in the two languages are completely different.
Such insufficient modeling on the transfer of local and global speaking styles will synthesize speeches without those particular emotions and intentions 
and substantially decrease the expressiveness of the dubbed speech.

In this paper,
to address these issues and improve speaking style transfer in automatic dubbing,
we propose a multiscale cross-lingual speaking style transfer framework,
which jointly models and optimizes the speaking style transfer for both dubbing directions between two languages with multitask learning.
The global and local speaking styles of the speeches in each language are extracted and utilized to predict the corresponding global and local speaking styles in the other language.
The predicted global and local speaking styles for the other language are then synthesized to speech by
a corresponding multiscale speaking style-enhanced FastSpeech 2 (hereafter, 
MST-FastSpeech 2
) \cite{ren2020fastspeech, li_inferring_2022}
for each language.
The source code of our work is also available on GitHub\footnote{https://github.com/thuhcsi/StyleDub}.

We evaluate the effectiveness of our approach using the mean squared error (MSE) between the predicted mel spectrogram and the ground-truth mel spectrogram, as well as the mean option score (MOS) on naturalness and speaking style transfer.
We employ an approach that also jointly models both dubbing directions but transfers only word duration instead of the global and local speaking styles as the baseline.
The experimental results indicate that %
the proposed approach outperforms the baseline approach
in objective and subjective evaluations.

\iftrue
The remainder of the paper is organized as follows:
We introduce related work in Section \ref{related}.
We then conduct data observation
in Section \ref{observation}.
In Section \ref{method}, we introduce the details of our proposed approach.
Experimental results and ablation studies are presented in Section \ref{experiments}.
We discuss our work in Section \ref{discussion}.
Last, %
Section \ref{conclusion}
concludes the paper.
\fi

\section{Related work}
\label{related}

\subsection{Text-to-speech synthesis}
Text-to-speech synthesis
is still a challenging task despite decades of investigations \cite{taylor2009text}.
Conventional TTS systems usually employ complicated pipelines\cite{zen2009statistical},
including linguistic feature generation, duration prediction, acoustic feature prediction and vocoder\cite{oord2016wavenet, kong2020hifi}. %
These components complete the respective subtasks in the conventional TTS pipelines
but are separately developed and trained, suffering from errors compounding in steps \cite{graves2014towards, wang2017tacotron}.

With the development of deep learning,
end-to-end TTS techniques are proposed to integrate the components in conventional TTS systems \cite{wang2017tacotron, shen2018natural} into one trainable framework.
End-to-end TTS systems, such as Tacotron \cite{wang2017tacotron, shen2018natural} and FastSpeech \cite{ren2020fastspeech},
are benefiting from joint training to alleviate the compounding errors and provide more robust synthesized speeches with less feature engineering.
These end-to-end TTS systems are powered by an encoder-decoder structure \cite{cho2014learning} with an attention mechanism \cite{bahdanau2014neural}, using an encoder to produce linguistic encodings and a decoder to directly generate raw spectrograms.
An attention mechanism such as location-sensitive attention in \cite{chorowski2015attention} is applied with the decoder to increase both performance and interpretability by
learning the alignment between the text and the spectrogram.
Then, a neural vocoder is utilized to generate waveforms from the spectrogram
to further improve the overall naturalness and quality of the synthesized speech \cite{shen2018natural}.

The improvement of end-to-end TTS systems
reveals a growing opportunity for many applications, such as audiobook narration, news readers, and conversational assistants,
which also challenges end-to-end TTS systems to improve the controllability of
choosing appropriate speaking styles for different application scenarios.
Global style token \cite{wang2018style} is proposed
to represent the %
attributes that are related to
the speaking style among the whole utterance,
such as emotion and intention,
as a single vector,
which
has been widely utilized in
robust emotion and global speaking style transfer
\cite{wang2018style}.
An adversarial branch \cite{ganin2016domain} is further proposed to disentangle
the speaker information from global style tokens
for cross-speaker global speaking style transfer \cite{wang2018style}.
At the word level,
local style token \cite{li_inferring_2022} is proposed to
represent the attributes related to the speaking style of each word, such as emphasis, rhythm and intonation
using a neural network-based force aligner \cite{li2022neufa}.

\subsection{Automatic dubbing}
Dubbing is a complex process of translating speeches in films or games to another language while preserving the original speaking style,
which can be regarded as a special kind of speech-to-speech translation\cite{jia2019direct, lee2021textless, kikui2003creating,brannon_dubbing_2023,chaume_dubbing_2020}.
Traditionally, such a dubbing process is performed by professional teams and voice talent,
in which voice talent will imitate pitch and energy and other speaking style-related properties in the original speech at the utterance and word levels \cite{brannon_dubbing_2023}.
Nevertheless, despite the significant amount of workforce engaged in it, some dubbing works are still controversial among audiences due to the unsatisfactory performance of the speaking style in the translated speeches.

With the development of TTS synthesis, neural network-based automatic dubbing is a possibility %
to be developed on the top of end-to-end TTS synthesis technologies to simplify and standardize such dubbing processes in film and game production.
Several automatic dubbing systems have been proposed to transfer the duration and speaking rate to dubbed speech. %
A prosodic alignment-based automatic dubbing system is proposed
\cite{oktem2019prosodic}
to transfer the duration of each prosodic phrase. %
This approach employs
a prosodic alignment algorithm to extract the mapping between the prosodic phrases in two languages from the attention weights extracted from a pretrained machine translation model.
The algorithm then adjusts the duration of each prosodic phrase to be the same as the duration of its mapped prosodic phrase in the original language.
However,
such a design could generate a duration that is too long or too short when
the number of phonemes between the mapped prosodic phrases in the two languages is largely different.

A HMM
\cite{rabiner1986introduction}-based prosodic alignment algorithm
\cite{federico2020speech} 
is further proposed to segment dubbed speech into prosodic phrases
that match the pauses in the original speech
by maximizing the probability of pause positions in the dubbed speech.
An improvement
\cite{virkar2021improvements}
to this HMM-based prosodic alignment algorithm is proposed to also transfer the speaking rate.
An additional phoneme-level duration predictor is proposed \cite{9747158}
to further improve the predicted duration and speaking rate
in both prosodic alignment and the following TTS system.
However, these approaches disregard the semantic meaning of each word during the segmentation of prosodic phrases. %
Thus, it is very likely that
the original and corresponding segmented prosodic phrases have completely different semantic meanings.
In this case, the duration and speaking rate in the original language is forcibly transferred to an unrelated prosodic phrase in the dubbed language,
which may substantially lower the naturalness of the synthesized speech.

\subsection{Contributions}
Compared with previous works, the contributions of our work are presented as follows:
(1) To the best of our knowledge, this is the first work that introduces multiscale speaking style transfer to automatic dubbing, in which we advocate the modeling of speaking style transfer at both the global scale and local scale to significantly improve expressiveness in automatic dubbing;
(2) This is also the first work that employs multitask learning to jointly optimize the modeling of cross-lingual speaking style transfer for both dubbing directions between two languages;
(3) We propose an effective approach to infer speaking styles at both the global scale and local scale for automatic dubbing.

\section{Data observation}
\label{observation}
We conduct subjective and objective observations to analyze the speaking style transfer among real-world professional dubs.
We randomly collect 25 pairs of parallel utterances in the English and Chinese dubs of the game \textit{Borderlands 3} for the observations,
which is a well-known game with excellent and highly acclaimed dubs in many languages.

\subsection{Subjective observation}
\label{subjective observation}
In the subjective observation,
25 listeners are invited to
listen to the English and Chinese speeches of each pair of these collected utterances and answer the following questions:
(1) Is there a \textit{special} global speaking style in each utterance that differs from the \textit{normal} speaking style in declarative sentences?
(2) Are the global speaking styles of this pair of utterances in the two languages the same, similar, irrelevant or opposite?
(3) Is there any \textit{special} local speaking style in the words of each utterance that differs from the \textit{normal} speaking style in the words of declarative sentences?
(4) If a word has a word in the other language that has the same or similar meaning and local speaking style as this word, we say that this word can be \textit{matched} to the corresponding word in the other language. Then, for each language, approximately how many words can be \textit{matched} to the words in another language?

The first two questions demonstrate the necessity and possibility of modeling global speaking style transfer in automatic dubbing.
The aggregated responses from the listeners (by averaging) indicate that $76.1\%$ and $78.1\%$
of the dubbed utterances in English and Chinese, respectively, have global speaking styles that are different from the speaking style in common declarative sentences.
This finding shows that various global speaking styles, 
which should be considered and transferred to other languages, are commonly employed in dubbing.
In addition,
the aggregated responses further indicate that in $96.8\%$ of the pairs of parallel utterances in these two languages,
their global speaking styles are the same or similar.
This finding demonstrates the possibility of modeling global speaking style transfer in automatic dubbing.

In contrast,
the  next two questions are designed to demonstrate the necessity and possibility of modeling speaking style transfer at the local scale.
The aggregated responses to the third question show that $83.1\%$ and $83.7\%$
of the dubbed utterances in English and Chinese, respectively, have words with local speaking styles different from the speaking style in common declarative sentences,
which reflects the importance of modeling local speaking style transfer for automatic dubbing.
In the final question,
the aggregated responses show that $86.7\%$ of the words in the English dubs can be \textit{matched}
to the words in Chinese, which means that they have similar semantic information and local speaking styles.
For the other direction, $87.1\%$ of the words in the Chinese dubs can also be \textit{matched} to the words in English.
This reveals that the similarity of the semantic meanings is related
to the similarity of local speaking styles between parallel utterances in different languages,
and shows the possibility of modeling local speaking style transfer across languages.

\iftrue
We summarize the aggregated responses in Table \ref{tab:observation1}.
The above observation demonstrates the need to consider global and local speaking styles in automatic dubbing (Questions 1 and 3),
and the possibility of modeling the speaking style transfer between the utterances at the global scale (Question 2)
and those at the local scale (Question 4).
\fi

\begin{figure*}[tb]
	\centering
	\includegraphics[width=.95\linewidth]{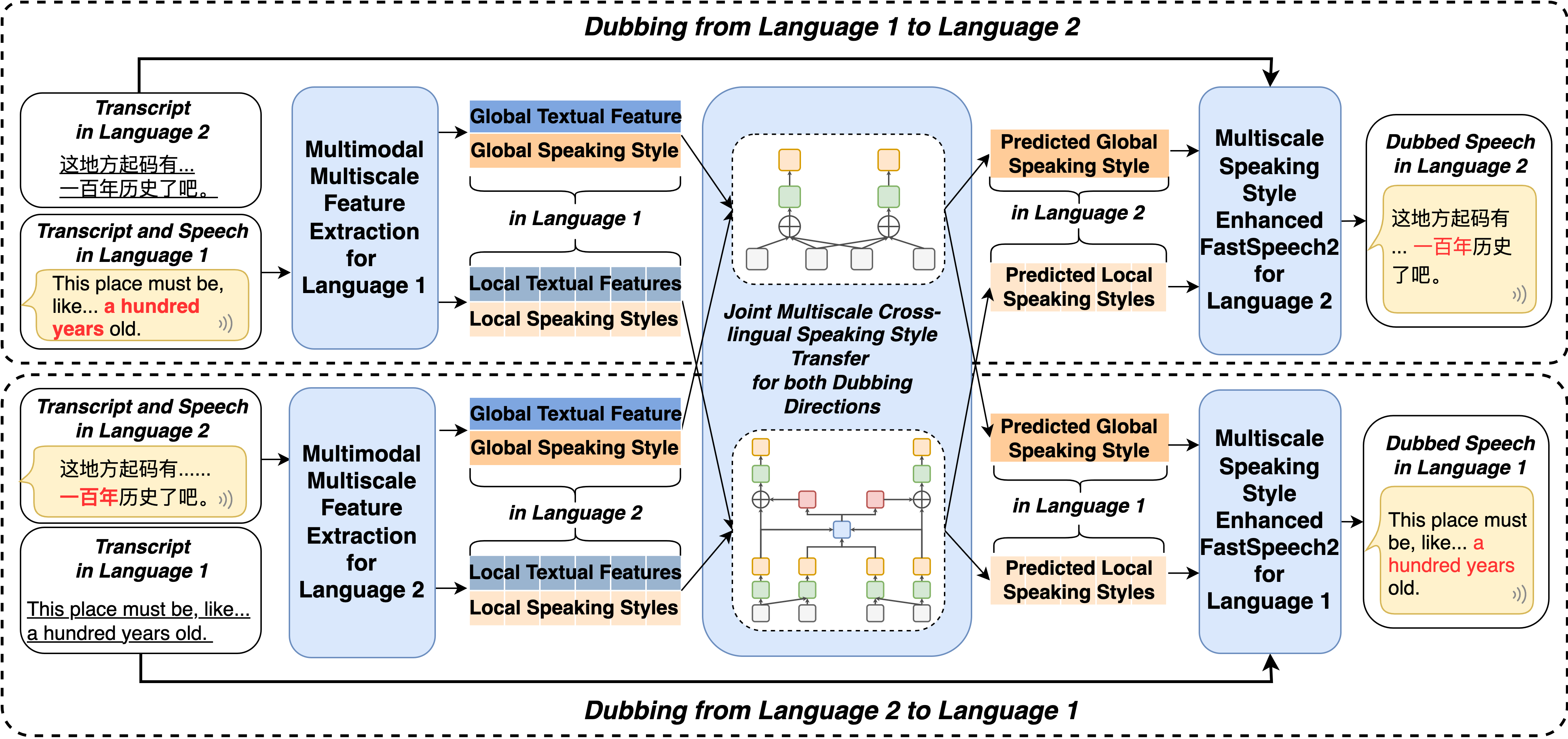}
	\caption{Architecture of the proposed joint multiscale cross-lingual speaking style transfer framework.}
	\label{fig:model}
\end{figure*}

\subsection{Objective observation}

\input{table/observation}

We refer to the experiments in a large-scale analysis on the human dubbing process \cite{brannon_dubbing_2023} to conduct an objective observation. %
The results are summarized in Table \ref{tab:observation2}.

First, we calculate the Pearson correlations of speaking rate and the means and standard deviations of pitch and energy between these pairs of utterances.
The results
suggest that these parallel dubs are correlated across these properties at the utterance level.
Second, we calculate the same Pearson correlations
between words that could be obviously \textit{matched} in the two languages.
The duration of each word is obtained by the Montreal Forced Aligner \cite{mcauliffe_montreal_2017}.
The results show lower Pearson correlations than those at the utterance level,
since the pronunciations of these words in English and Chinese are completely different.
However, the \textit{matched} words are still positively correlated with each other across these properties.

The patterns identified at both the utterance level and word level are similar to those in the large-scale analysis \cite{brannon_dubbing_2023}.
The patterns indicate that human dubbers imitate the properties of the source audio at both the global scale and local scale,
which also demonstrates the need for multiscale speaking style transfer in automatic dubbing.

\section{Methodology}
\label{method}
Based on the
data observation,
we propose a multiscale cross-lingual speaking style transfer framework for automatic dubbing,
as illustrated in Figure \ref{fig:model},
which jointly learns the speaking style transfer between two languages at both the global scale and local scale.

\subsection{Multimodal multiscale feature extraction}
To support the modeling
of multiscale cross-lingual speaking style transfer,
the textual features and multimodal features that fuse the textual and speaking style information in the two languages are extracted at both the global scale and local scale.

\subsubsection{Textural feature extraction}
A pretrained bidirectional encoder representations from transformers (BERT) \cite{devlin_bert_2019} model is employed for each language to extract
the sentence-level and word-level BERT embeddings.
The sentence-level BERT embedding is directly applied as the textual feature at the global scale for each language.
The word-level BERT embeddings are converted by a local text encoder for each language as the textual features at the local scale.
The local text encoder consists of the pre-net and CBHG networks in Tacotron \cite{wang2017tacotron}.
The above process can be formulated as:
\begin{align}
    g_1^t &= SBERT_1\\
    g_2^t &= SBERT_2\\
    s_1^t &= E_1^t\left(BERT_1\right)\\
    s_2^t &= E_2^t\left(BERT_2\right)
\end{align}
where
$SBERT_1$, $SBERT_2$, $g_1^t$ and $g_2^t \in \mathcal{R}^{d_{b}}$
are the sentence-level BERT embeddings
and global textual features
for each language,
$BERT_1 \in \mathcal{R}^{l_1\times d_b}$ and $BERT_2 \in \mathcal{R}^{l_2\times d_b}$
are the word-level BERT embeddings for each language,
$d_b$ is the dimension of sentence- and word-level BERT embeddings,
$l_1$ and $l_2$ are the numbers of words in the two utterances,
$E_1^t$ and $E_2^t$ are the local text encoders,
$s_2^t \in \mathcal{R}^{l_1\times d_e}$ and $s_1^t\in\mathcal{R}^{l_2\times d_e}$ are the local textual features for each language, and
$d_e$ is the output dimension of the local text encoder.

\subsubsection{Speaking style extraction}
Global and local speaking style encoders \cite{li_inferring_2022} are then employed to extract the speaking styles at global and local scales for each language.
The global speaking style encoder consists of 6 strided convolutional neural networks (CNNs) composed of $3\times3$ kernels
with $2\times2$ stride and 32, 32, 64, 64, 128, 128 filters, respectively,
a 256-dimensional GRU layer and a 128-dimensional style attention layer.
The mel spectrograms of the input speech are processed by the CNNs and GRU.
The final state of the GRU is further sent to the style attention layer to derive the global speaking style vector $GST$ as
the weights for 10 automatically learned base global speaking style embeddings.
The process can
be formulated as:
\begin{align}
    q &= final(GRU(CNN(mel))) \\
    GST &= softmax\left(q^TGST^{table}\right)
\end{align}
where %
$final$ returns the final state of the GRU,
$q$ is the query for the style attention layer, and
$GST^{table}$ contains the 10 automatically learned base global speaking style embeddings.

The architecture of the local speaking style encoder is the same as the global speaking style encoder,
with the exception that the stride is now %
$1\times2$ and the GRU layer now returns the output for each input frame.
The outputs of the GRU are then summarized for each word in the utterance by multiplying them with the speech-to-text attention weights extracted from a pretrained neural network-based forced aligner (NeuFA) \cite{li2022neufa}.
The NeuFA
employs a bidirectional attention mechanism \cite{li2022neufa} to learn the bidirectional information mapping between a pair of text and speech.
The learned attention weights in the speech-to-text direction can be used to summarize the
frame-level information
for each word in the utterance, deriving the local speaking style sequence $LST$.
The process of the local reference encoder
can be formulated as:
\begin{align}
    q' &= W_{ASR}^TGRU(CNN(mel)) \\
    LST &= softmax\left(q'^TLST^{table}\right)
\end{align}
where
$W_{ASR}$ is the attention weight in the speech-to-text direction obtained by NeuFA,
$q'$ is the query for the local style attention layer, and
$LST^{table}$ contains 10 automatically learned base local speaking style embeddings.

\subsubsection{Multimodal feature fusion}
To fuse the extracted textual and features, %
the sentence-level BERT embedding and the global speaking style of each language are then concatenated as the multimodal feature at the global scale.
The word-level BERT embeddings and local speaking styles are concatenated and consumed by a local encoder for each language to produce the multimodal feature sequence at the local scale.
Here, the local encoder has the same architecture as the local textual encoder.
The process can be formulated as:
\begin{align}
    g_1^m &= \left[SBERT_1; GST_1\right]\\
    g_2^m &= \left[SBERT_2; GST_2\right]\\
    \label{local feature 1}
    s_1^m &= E_1\left(\left[BERT_1; LST_1\right]\right)\\
    \label{local feature 2}
    s_2^m &= E_2\left(\left[BERT_2; LST_2\right]\right)
\end{align}
where $GST_1$, $GST_{2} \in\mathcal{R}^{d_{st}}$ are the global speaking style for each language,
$LST_1\in\mathcal{R}^{l_1\times d_{st}}$ and $LST_2\in\mathcal{R}^{l_2\times d_{st}}$ are the local speaking styles for each language,
$E_1$ and $E_2$,
$g_1^m\in\mathcal{R}^{d_b+d_{st}}$ and $g_2^m\in\mathcal{R}^{d_b+d_{st}}$, $s_1^m\in\mathcal{R}^{l_1\times d_{e}}$ and $s_2^m\in\mathcal{R}^{l_2\times d_{e}}$
are local encoders,
global multimodal features
and local multimodal features,
for each language.

\subsection{Joint multiscale cross-lingual speaking style transfer}%

\begin{figure}[t]
	\centering
	\includegraphics[width=\linewidth]{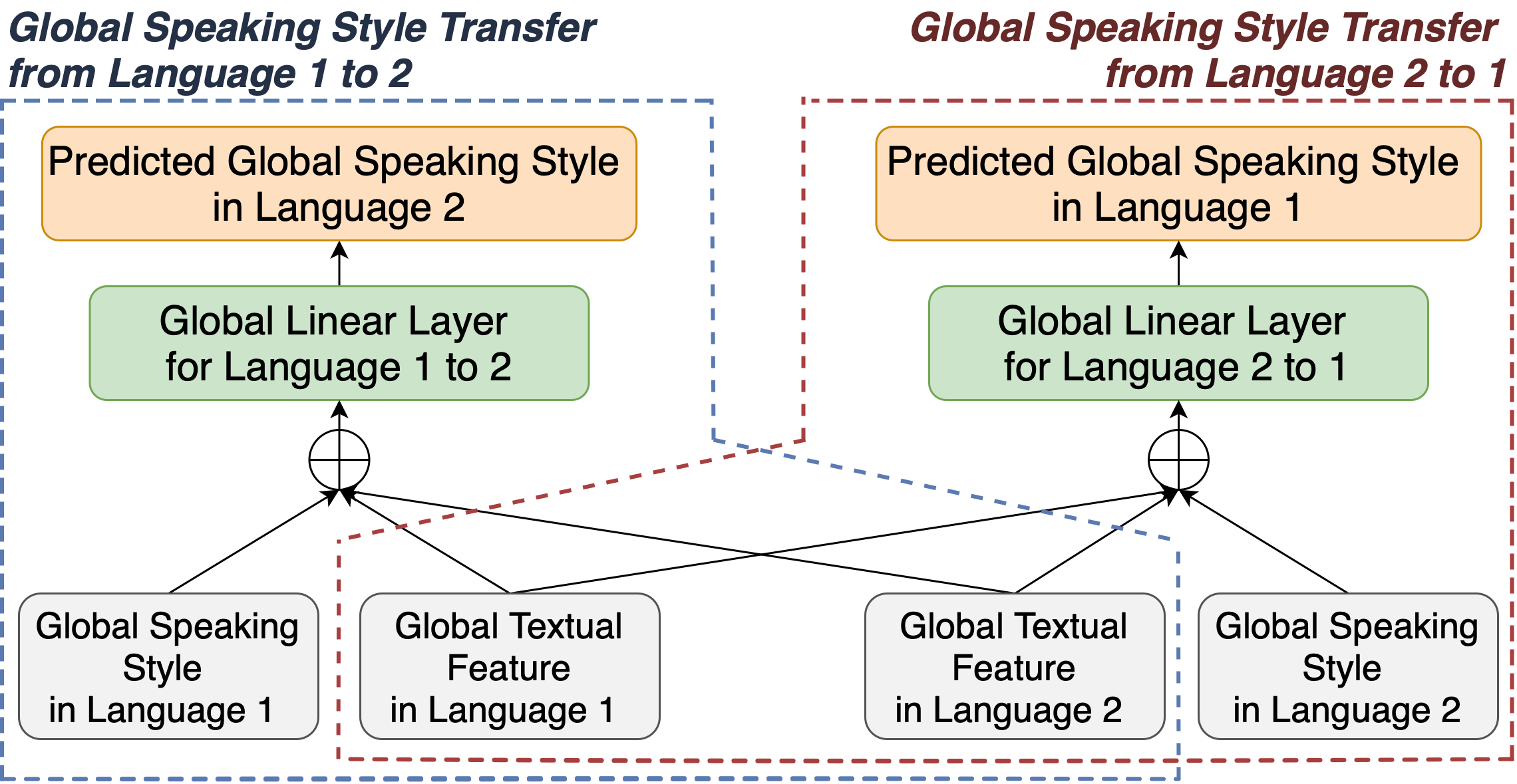}
	\caption{Joint cross-lingual speaking style transfer at the global scale.}
	\label{fig:global}
\end{figure}

\subsubsection{Speaking style transfer at the global scale}

As shown in Figure \ref{fig:global},
the global multimodal feature of each language
is concatenated
to the global textual feature of the other language
and is consumed by a global linear projection layer to predict the global speaking style in the other language:
\begin{align}
    GST'_1 &= f_{2to1}^g([g_2^m; g_1^t])\\
    GST'_2 &= f_{1to2}^g([g_1^m; g_2^t])
\end{align}
where
$f_{2to1}^g$ and $f_{1to2}^g$,
$GST'_1$ and $GST'_2$
are the global linear projection layer for each dubbing direction and the predicted global speaking style for each language.

\begin{figure*}[t]
	\centering
	\includegraphics[width=.9\linewidth]{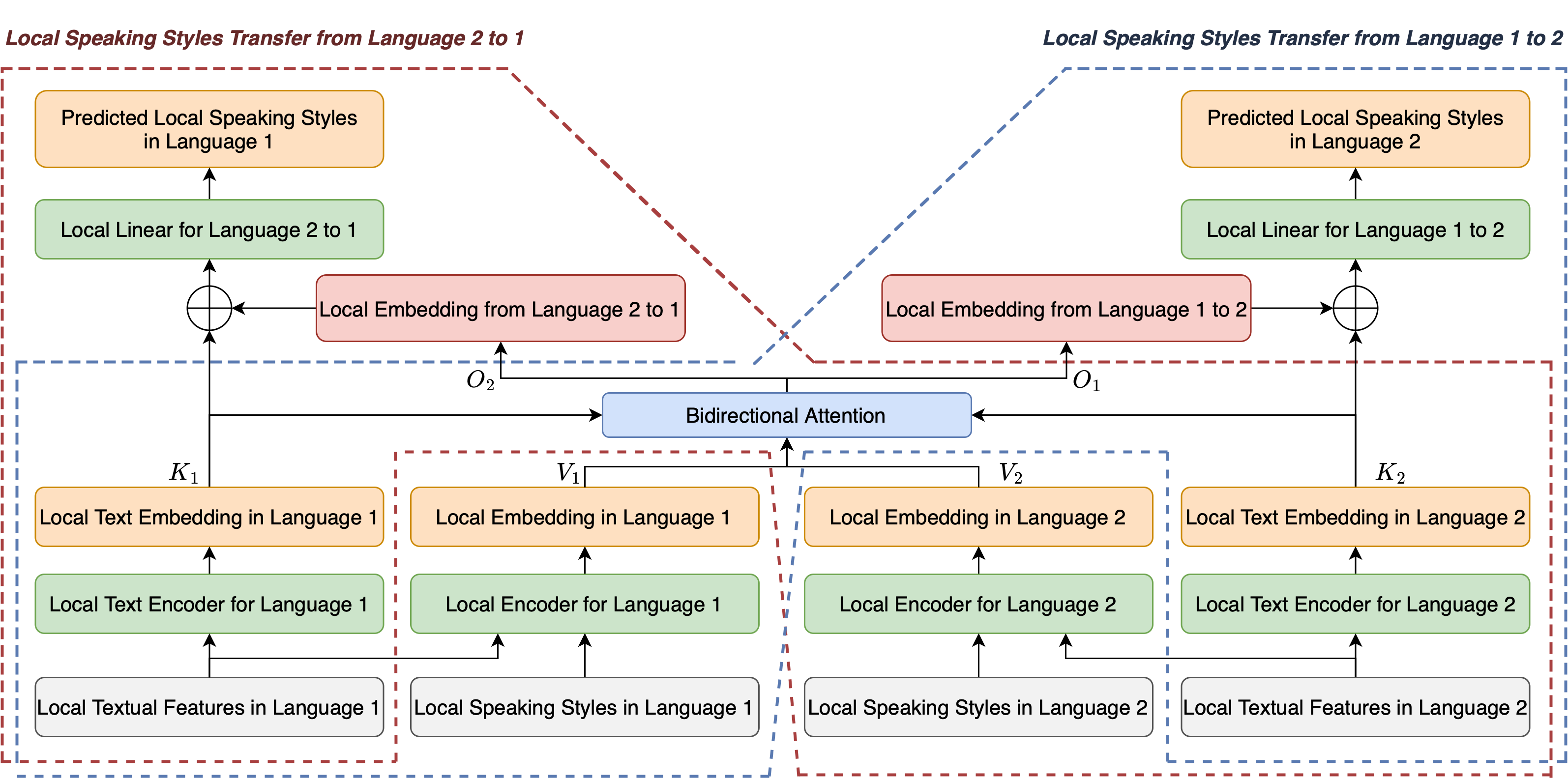}
	\caption{Joint cross-lingual speaking style transfer at the local scale.}
	\label{fig:local}
\end{figure*}

\begin{figure}[b]
	\centering
	\includegraphics[width=\linewidth]{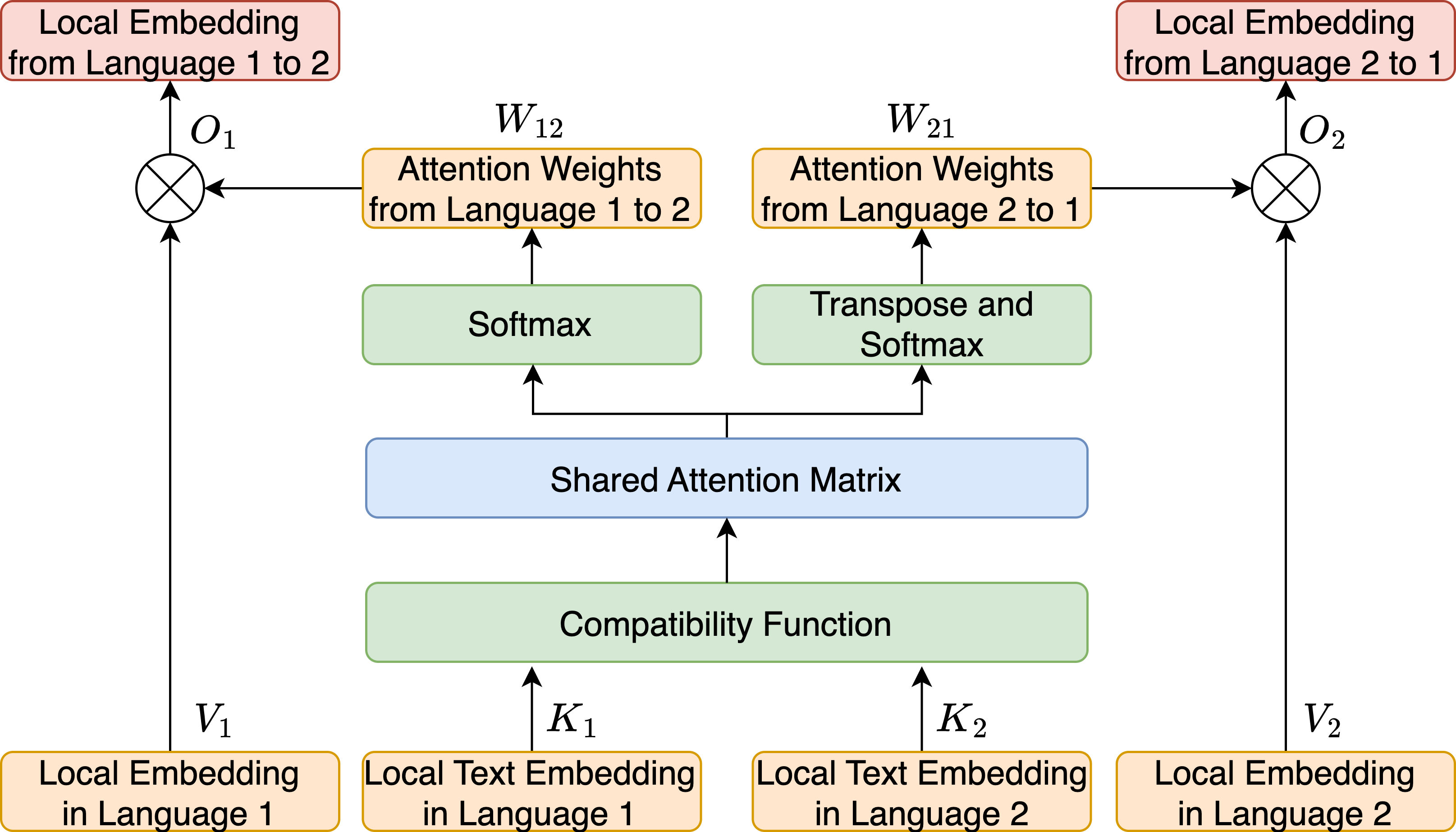}
	\caption{
	    Bidirectional attention mechanism in local speaking style transfer.
	}
	\label{fig:attention}
\end{figure}

\subsubsection{Speaking style transfer at the local scale}

As shown in Figure \ref{fig:local},
an architecture with two encoder-decoder frameworks and a shared bidirectional attention mechanism
is employed to learn the speaking style transfer between two languages at the local scale.

The bidirectional attention mechanism is extended from the conventional attention mechanism \cite{vaswani2017attention} 
to establish the bidirectional relationship between two sets of key-value pairs, as illustrated in Figure \ref{fig:attention}
\cite{li2022neufa}.
The bidirectional attention mechanism uses two sets of key-value pairs as inputs, whereby each set's keys function as the queries for the other set's key-value pairs.
Subsequently, a shared compatibility function is modeled for both sets of keys to calculate the attention weights for each direction.
The attention weights are then utilized to summarize the two sets of values as the outputs for both directions.
The bidirectional attention mechanism is formulated as:
\begin{align}
    A &= f(K_1, K_2) \\
    \label{bidirectional attention2}
    W_{12} &= softmax(A)\\
    W_{21} &= softmax(A^T)\\
    O_1 &= W_{12}^{T}V_1\\
    O_2 &= W_{21}^{T}V_2
\end{align}
where $K_1 \in R^{n_1 \times d_{k1}}$, $V_1 \in R^{n_1 \times d_{v1}}$ and
$K_2 \in R^{n_2 \times d_{k2}}$, $V_2 \in R^{n_2 \times d_{v2}}$
are the two sets of key-value pairs;
$n_1$ and $n_2$ are the numbers of the key-value pairs; %
$d_{k1}$, $d_{v1}$, $d_{k2}$ and $d_{v2}$ are the feature dimensions; %
$f$ is the compatibility function;
$A \in R^{n_1 \times n_2}$ is the shared attention matrix;
$W_{12} \in R^{n_1 \times n_2}$ and
$W_{21} \in R^{n_2 \times n_1}$
are the attention weights for two directions;
$O_1 \in R^{n_2 \times d_{v1}}$ is the weighted sum of $V_1$ for each key in $K_2$; and
$O_2 \in R^{n_1 \times d_{v2}}$ is the weighted sum of $V_2$ for each key in $K_1$.

We employ the multiplicative form of the bidirectional attention mechanism to jointly model speaking style transfer at the local scale
and summarize the multimodal features for both dubbing directions.
In particular, the local textual features and local multimodal features of the two languages are employed as the two sets of keys
and values, and the calculations of the attention weights and outputs are formulated as:
\begin{align}
    A_{1to2} &= f_1(s_1^t) \times f_2(s_2^t)^T\\
    A_{2to1} &= f_2(s_2^t) \times f_1(s_1^t)^T = A_{1to2}^T\\
    W_{1to2} &= softmax(A_{1to2})\\
    W_{2to1} &= softmax(A_{2to1})\\
    O_{1to2} &= W_{1to2}^Ts_1^m\\
    O_{2to1} &= W_{2to1}^Ts_2^m
\end{align}
where
$f_1$ and $f_2$ are two linear projections,
$A_{1to2}$ and $A_{2to1}$ are the shared attention matrices,
$W_{1to2}$ and $W_{2to1}$ are the learned attention weights for the two dubbing directions, and
$O_{1to2}$ and $O_{2to1}$ are the summarized multimodal features of each language. %

The summarized multimodal features of each language
are then concatenated with the local textual features of the other language
and consumed by a local linear projection to the predicted local speaking styles of the other language:
\begin{align}
    \label{lst predict 1}
    LST'_1 &= f_{2to1}^l([O_{2to1}; s_1^t])\\
    \label{lst predict 2}
    LST'_2 &= f_{1to2}^l([O_{1to2}; s_2^t])
\end{align}
where
$f_{2to1}^l$ and $f_{1to2}^l$,
$LST'_1$ and $LST'_2$
are the local linear projections for each dubbing direction and the predicted local speaking styles for each language.

\subsection{Speech synthesis with predicted multiscale speaking styles}

An MST-FastSpeech 2 \cite{li_inferring_2022} is trained for each language to synthesize speeches with global and local speaking styles
inferred by the proposed framework.

The speaker embedding and the predicted global and local speaking styles of each utterance are upsampled to the phoneme level and concatenated with the encoder outputs of FastSpeech 2.
Then, the pitch, duration and energy of each phoneme
are inferred from the concatenated encoder outputs
by the variance adaptor,
and further converted to the mel spectrogram by the decoder of FastSpeech 2.

A pretrained HiFi-GAN \cite{kong2020hifi} then serves as the vocoder to synthesize speech from the predicted mel sprectrogram with desired speaking styles transferred from the other language.

\begin{figure}[tb]
	\centering
	\includegraphics[width=\linewidth]{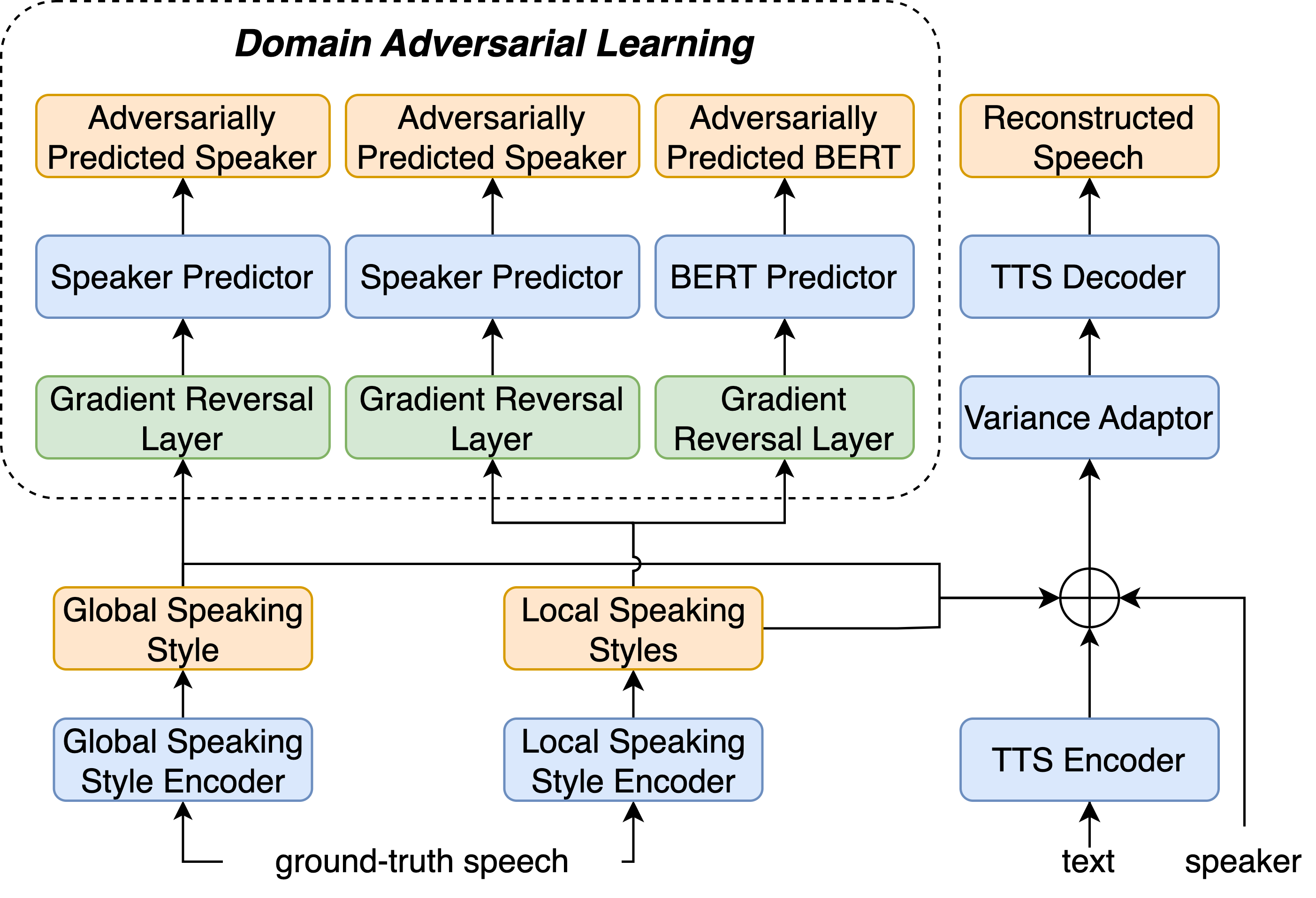}
	\caption{Pretraining the global and local speaking style encoders and FastSpeech 2 for each language\cite{li_inferring_2022}.}
	\label{fig:mst}
\end{figure}

\subsection{Training strategy}

Before the proposed joint cross-lingual multiscale speaking style transfer framework is trained,
the global and local speaking style encoders and their corresponding MST-FastSpeech 2 for each language
are trained
to ensure that the extracted global and local speaking styles are compatible with the TTS backbone.
To improve the robustness of the extracted speaking styles
and their cross-lingual transfer,
the text and speaker information
are also disentangled
from the global and local speaking styles
with
the gradient reversal layer (GRL) \cite{ganin2015unsupervised} and
domain adversarial learning \cite{ganin2016domain, zhang_learning_2019} at this stage:
\begin{align}
    speaker'_{GST} &= f_{speaker}^{GST}\left(GRL(GST)\right)\\
    speaker'_{LST} &= f_{speaker}^{LST}\left(GRL(LST)\right)\\
    BERT' &= f_{BERT}\left(GRL(LST)\right)
\end{align}
where $f_{speaker}^{GST}$, $f_{speaker}^{LST}$ and $f_{BERT}$ are linear projections
serving as the adversarial speaker predictors and text predictor,
GRL reverses the gradients of these adversarial predictors,
$speaker'_{GST}$ and $speaker'_{LST}$ are the adversarially predicted
speaker embeddings, and
$BERT'$ is adversarially predicted
textual embedding
of this utterance.
The loss for pretraining is the sum of the reconstruction loss and adversarial losses,
where the former is
the mean squared error (MSE) between the predicted mel spectrogram and the ground-truth mel spectrogram,
and the latter are
the MSEs between the predicted speaker and BERT embeddings and the ground-truth speaker and BERT embeddings.

The pretrained global and local encoders and FastSpeech 2 of each language are then frozen and utilized to extract the speaking styles
and
to synthesize speech with the predicted speaking styles
in
the proposed framework.
The proposed joint multiscale cross-lingual speaking style transfer framework is then trained and backpropagated
with the sum of the MSEs between the predicted global and local speaking styles and the ground-truth 
global and local speaking styles 
on the two languages:
\begin{align}
    \nonumber
    loss = &MSE(GST_1, GST'_1) + MSE(LST_1, LST'_1)\\
    & + MSE(GST_2, GST'_2) + MSE(LST_2, LST'_2)
\end{align}

\section{Experiments}
\label{experiments}

\subsection{Baselines}

To demonstrate the effectiveness of the proposed joint multiscale cross-lingual speaking style transfer framework,
we employ 2 approaches with different speaking style transfer methods as the baselines, which also employ FastSpeech 2 as the TTS backbone.

\subsubsection{No speaking style transfer}

The first baseline approach is a vanilla FastSpeech 2 \cite{ren2020fastspeech} with no speaking style transfer,
which is also representative of state-of-the-art conventional TTS and speech-to-speech translation systems with ideal translated scripts.

\subsubsection{Duration transfer}

Based on the proposed approach,
we implement another approach with only duration transfer as the second baseline.
This approach is inspired by a state-of-the-art automatic dubbing system \cite{9747158} and should
have similar or better performance
with encoder-decoder frameworks and attention mechanisms.
In this approach, the modeling of global speaking style transfer is omitted,
and only the duration of each word is transferred to speech in the other language.
Specifically, a sequence of the duration for each word is utilized to replace the local speaking style in the proposed approach.
Equations (\ref{local feature 1}), (\ref{local feature 2}), (\ref{lst predict 1}) and (\ref{lst predict 2}) are changed to:
\begin{align}
    s_1^m &= E_1^l\left(\left[BERT_1; d_1\right]\right)\\
    s_2^m &= E_2^l\left(\left[BERT_2; d_2\right]\right)\\
    d'_1 &= f_{2to1}^l([O_{2to1}; s_1^t])\\
    d'_2 &= f_{2to1}^l([O_{1to2}; s_2^t])
\end{align}
where $d_1$, $d_2$, $d'_1$, and $d'_2$ are the sequences of duration and predicted duration for each word in each language.
Then, the sequence of predicted duration for each language is utilized by FastSpeech 2 with precise duration control for each language to synthesize the dubbed speech.

\subsection{Training setup}

We collect more parallel utterances from the English and Chinese dubs of the game \textit{Borderlands 3}
as the corpus for our research.
As introduced in Section \ref{observation}, the game \textit{Borderlands 3} is an action role-playing, first-person shooter video game
from the \textit{Borderlands} game franchise in a space Western science fiction setting,
which provides excellent and highly acclaimed dubs in many languages with various speaking styles.
\textit{Borderlands 3} has 4 playable characters and dozens of nonplayable characters (NPCs) in which there are approximately 20 main NPCs
who frequently speak with each other to tell background stories or guide the players in the missions.
Each of these playable and nonplayable characters has a unique speaking style to present a different personality.
Moreover,
the characters are further changing their speaking styles
for different scenarios during the 35 gameplay hours of main and other side missions.

We collect 4,914 pairs of parallel utterances from \textit{Borderlands 3},
including 5 hours of English speeches, 5 hours of Chinese speeches and their corresponding subtitles.
The sample rate of all the speeches in the corpus is 48,000 Hz.
For the English speeches in the corpus,
the length of each utterance ranged from 0.21 s to 12.63 s, with a mean of 3.80 s and a standard deviation of 2.35 s.
The number of words of each utterance ranged from 1 to 46, with a mean of 11.40 and a standard deviation of 7.72.
For the Chinese speeches in the corpus,
the length of each utterance ranged from 0.31 s to 12.63 s, with a mean of 3.91 s and a standard deviation of 2.44 s.
The number of words of each utterance ranged from 1 to 70, with a mean of 17.36 and a standard deviation of 11.92.

We extract mel spectrogram for each utterance in the corpus with a window length of 25 ms and a shift of 10 ms,
which will serve as the inputs for the global and local speaking style encoders and the ground truths for MST-FastSpeech 2.
We employ a pretrained BERT \cite{devlin2018bert} model to extract the sentence and word level BERT embeddings,
in which the BERT embedding of each word is the average of the BERT embeddings of the corresponding subwords.

We then employ the whole corpus to pretrain MST-FastSpeech 2 and the global and local speaking style encoders for both languages.
We follow the training setups of FastSpeech 2 \cite{ren2020fastspeech}
to train the model of each language for 500,000 iterations with a batch size of 32.

We randomly employ 4,414 pairs of parallel utterances in the two languages as the training set
to train the proposed multiscale speaking style transfer framework.
The proposed framework is trained for 10 epochs with a batch size of 32 and a learning rate of $10^{-4}$.

All the models are implemented with PyTorch \cite{paszke2019pytorch} and trained on an NVIDIA Tesla V100 GPU.

\input{table/results2}

\subsection{Evaluation}

We employ the remaining 500 pairs of utterances as the test set for evaluation.
We adopt the mean squared error (MSE) between the predicted mel spectrogram and the ground-truth mel spectrogram of the full band
as the metrics for the objective evaluation.
We also calculate %
the MSEs between the lowest 10 dimensions and the highest 10 dimensions of the mel spectrograms
to investigate
how the speaking styles are transferred at the global and local scales.
The predicted mel spectrogram is resized to match the length of the ground-truth mel spectrogram with nearest-neighbor interpolation.
We adopt nearest-neighbor interpolation instead of dynamic time wrapping (DTW) since DTW significantly affects the duration of each word, which is also an important attribute in speaking style.
Nearest-neighbor interpolation, on the other hand, does not have a major impact on the duration because the duration of the entire sentence in each approach is always close.

For the subjective evaluation, 20 conversation chunks\footnote{Synthesized samples at https://thuhcsi.github.io/StyleDub/.}
are further
randomly selected
and evaluated by 25 listeners.
Listeners are asked to rate the naturalness and how the speaking styles of synthesized speeches %
are transferred from speech in the other language
on a scale from 1 to 5 with 1-point intervals,
from which subjective mean opinion scores (MOS) are calculated.
The listeners are asked to choose a preferred dubbed speech from the speeches generated by
different approaches, from which preference rates are calculated.

\begin{figure*}[tb]
	\centering  
	\subfigure[Input English speech]{\includegraphics[width=0.5\linewidth]{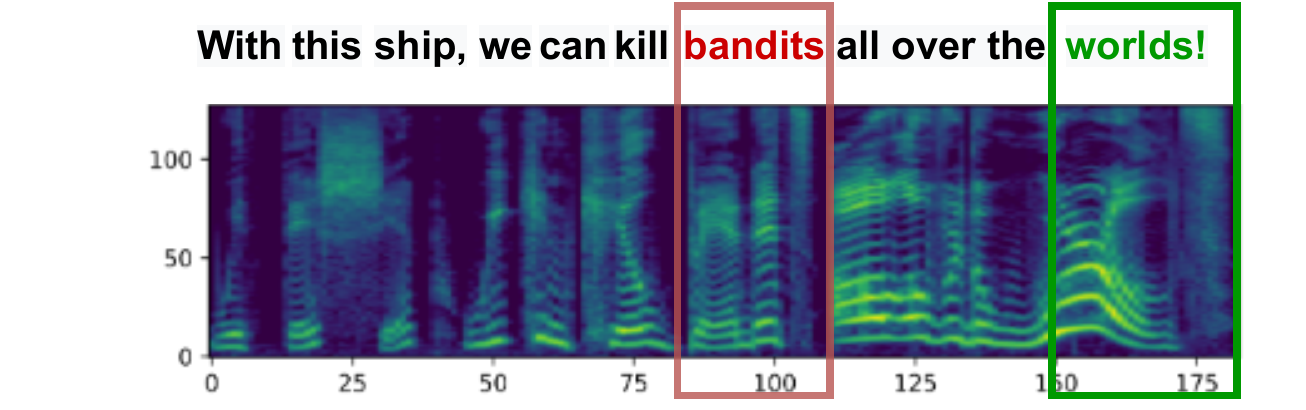}}%
	\subfigure[Input Chinese speech]{\includegraphics[width=0.5\linewidth]{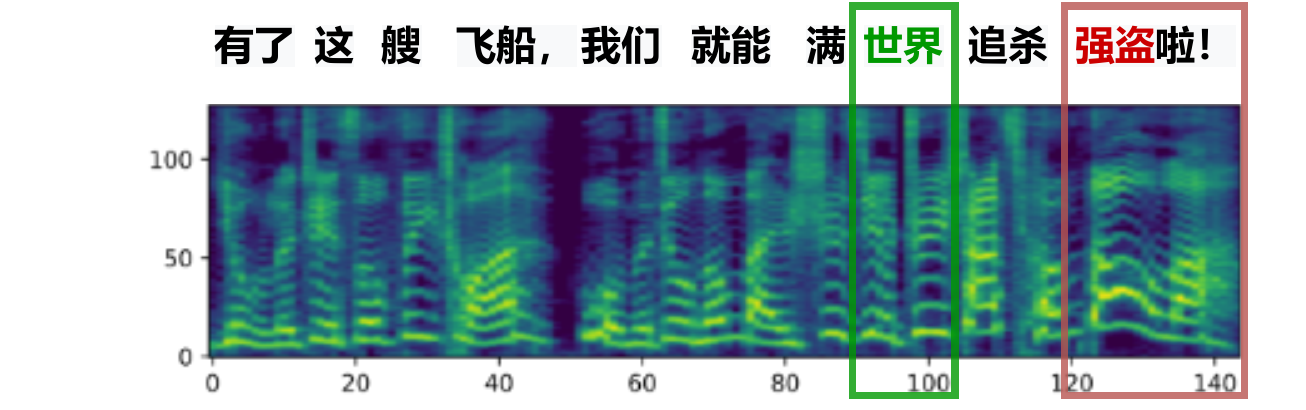}}\\
	\subfigure[Vanilla FastSpeech 2]{\includegraphics[width=0.5\linewidth]{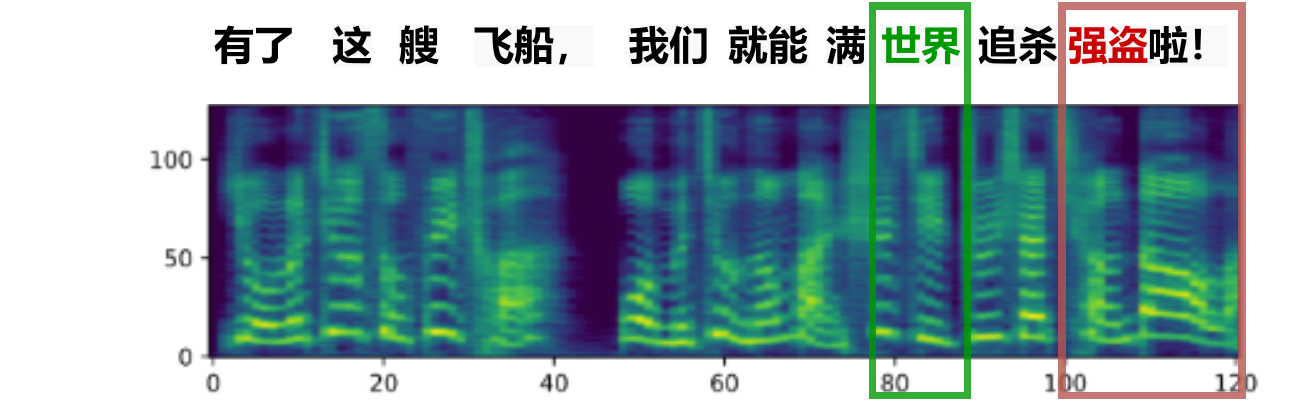}}%
	\subfigure[Vanilla FastSpeech 2]{\includegraphics[width=0.5\linewidth]{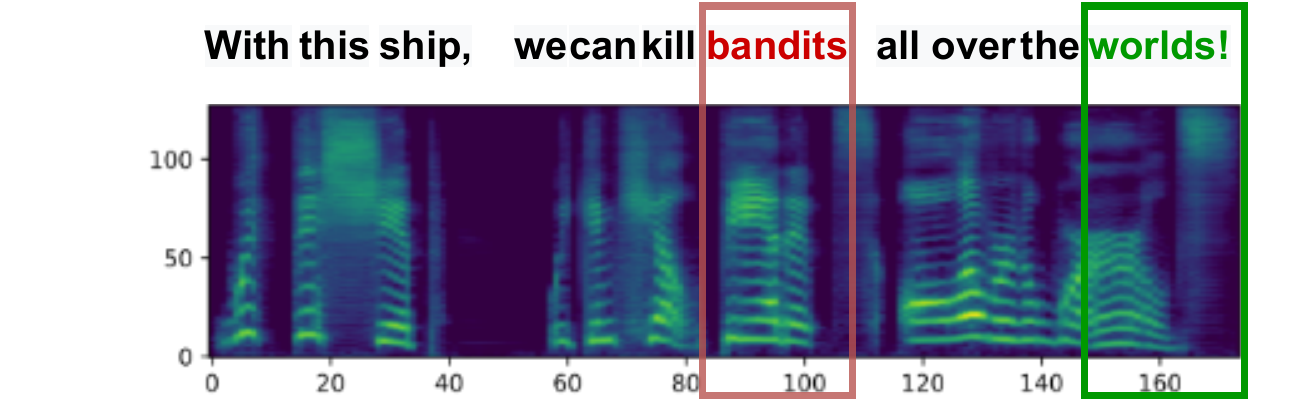}}\\
	\subfigure[Duration transfer]{\includegraphics[width=0.5\linewidth]{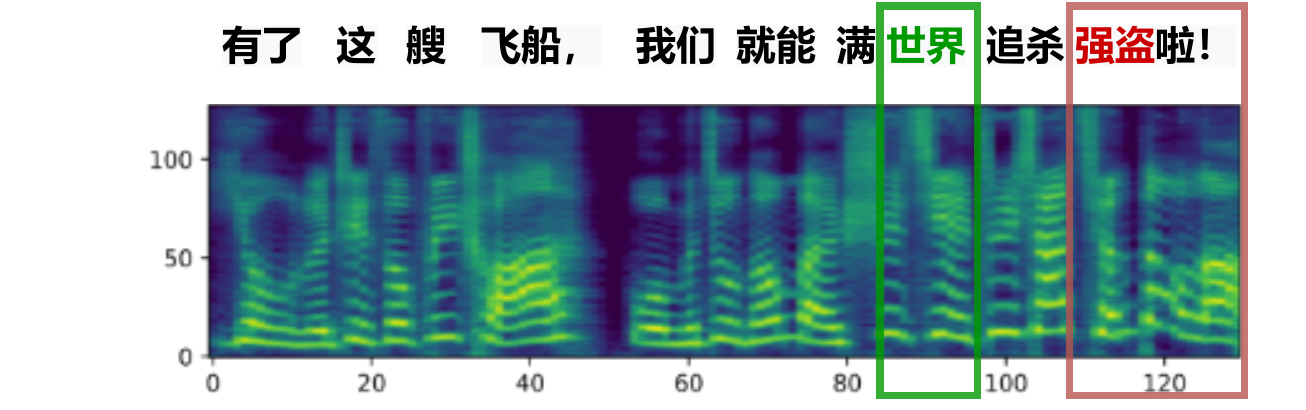}}%
	\subfigure[Duration transfer]{\includegraphics[width=0.5\linewidth]{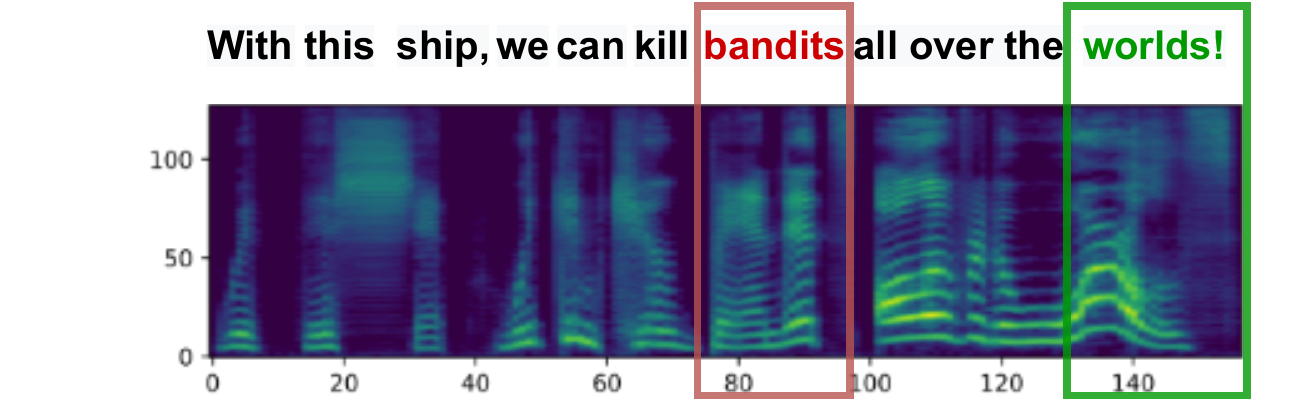}}\\
	\subfigure[Multi-scale speaking style transfer (Proposed)]{\includegraphics[width=0.5\linewidth]{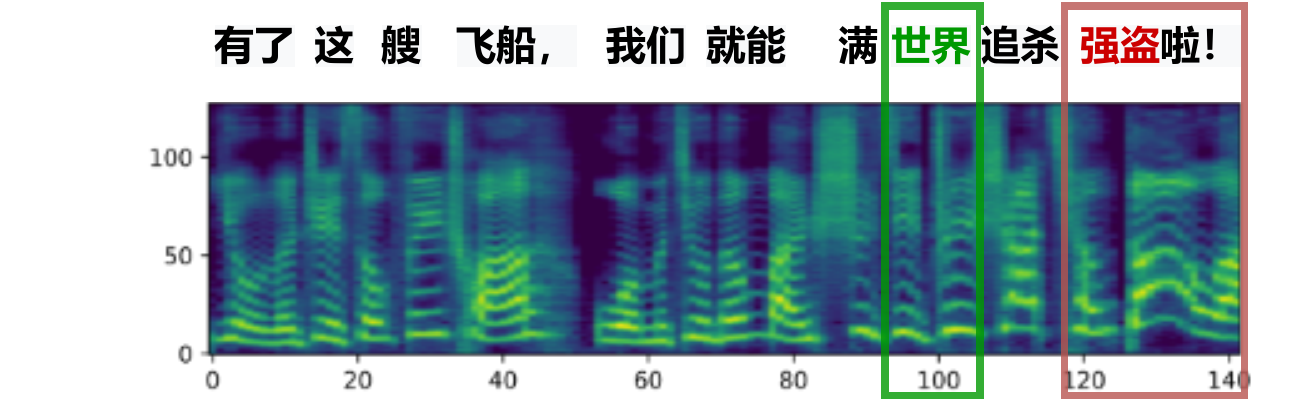}}%
	\subfigure[Multi-scale speaking style transfer (Proposed)]{\includegraphics[width=0.5\linewidth]{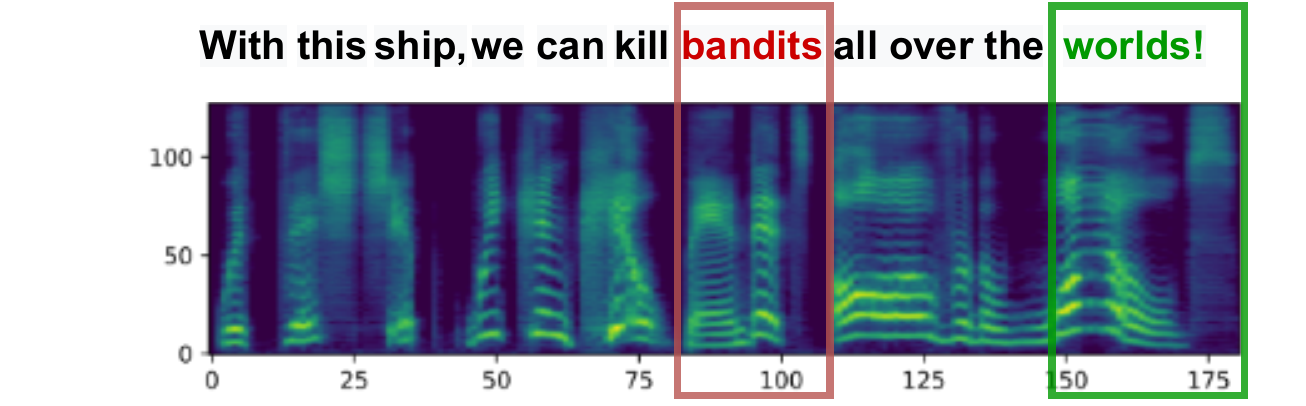}}\\
	\caption{Comparison between the mel spectrograms for the words `bandits' and `worlds' in the synthesized speech `With this ship, we can kill bandits all over the world' for each approach in English to Chinese (left column) and Chinese to English (right column) dubbing directions.}
	\label{fig:case}
\end{figure*}

\subsection{Experimental results}
The results of the objective and subjective evaluations are shown in Tables \ref{tab:results2} and \ref{tab:results}.
In the objective evaluations, both the approaches with duration transfer and the proposed multiscale speaking style transfer
outperform the baseline with no speaking style transfer.
And the proposed approach further significantly outperforms the baseline with duration transfer in the MSE of mel spectrogram at full, high and low frequency bands.
In the subjective evaluations,
the approach with duration transfer increases
the MOS on speaking style transfer by $0.762$ and $0.634$ for the two directions compared to the baseline with no speaking style transfer,
and the corresponding preference rate is increased by $18.35\%$ and $25.77\%$.
Compared to the baseline with duration transfer,
the proposed approach further improves the
MOS on speaking style transfer by $0.2$ and $0.013$ for the two directions, and the corresponding preference rate is increased by $40.05\%$ and $3.61\%$.
Such improvements in performance demonstrate the necessity and effectiveness of modeling the transfer on all aspects of speaking style,
not just duration.

Note that while the proposed model is found to deliver significantly better results than the baseline approach with duration transfer in the English to Chinese dubbing direction, the improvements are marginal when dubbing from Chinese to English.
This disparity may be explained by the lower density of speaking styles in Chinese dubs compared to English dubs.
During our subjective observation, most listeners report that the English dubs are more expressive and the speaking styles are more varied than the Chinese dubs.
As the English dubs are the original dubs for the game, there may have been a loss of speaking style during the dubbing process in other languages.
Additionally, voice actors commonly employ various accents for different characters when creating English dubs, whereas Chinese dubs are always dubbed in standard Mandarin without any accent.
This discrepancy may also cause difficulty when mapping the speaking styles in Chinese to English.

We illustrate the mel spectrogram of the synthesized speech by each approach in Figure \ref{fig:case} for the utterance
`With this ship, we can kill bandits all over the world' from the test set.
We highlight the comparison between the words `bandits' and `world' in the blue and red frames of these utterances.
In Figure \ref{fig:case} (a) and (b), these words in the original English and Chinese dubs are emphasized with local speaking styles to express the excitement of the characters with the new ship.
As shown in Figure \ref{fig:case} (g) and (h), their corresponding words in the other language
have also been more clearly emphasized than those synthesized by vanilla FastSpeech 2 and the baseline approach with only duration transfer.

\subsection{Ablation study}

Based on the proposed approach, we implement several variant approaches for ablation studies.
The results are shown in Table \ref{tab:ablation}.
In addition to the MSEs between the predicted mel spectrogram and the ground-truth mel spectrogram in both languages,
we further consider the MSEs between the predicted global and local speaking styles and the ground-truth global and local speaking styles in the ablation studies.

The first two variants disregard speaking style transfer at the global or local scale.
The TTS backbones of these variants have also excluded the modeling of global or local speaking styles.
These two approaches suffer from missing scales in speaking style transfer,
which yields a higher MSE of mel spectrograms than the proposed approach.
In particular, the approach that only transfers the local speaking styles outperforms the approach that only transfers the global speaking styles.
This finding validates the intuition that local speaking styles have greater importance in automatic dubbing.
However, solely transferring local speaking styles cannot account for sense-for-sense translations and words that cannot be \textit{matched} in another language.
The proposed approach,
benefitting from the joint modeling on both global and local speaking style transfer,
utilized local speaking style for \textit{matched} words and global speaking style for sense-for-sense translations, resulting in the lowest MSE of mel spectrograms.
Note that even though these two variants have a lower MSE of global or local speaking styles than the proposed approach, they cannot be compared because they are extracted from different TTS backbones.

The last two variants disregard one dubbing direction each during speaking style transfer.
The proposed approach slightly outperforms these two variants, 
indicating
 the effectiveness of jointly modeling speaking style transfer in both dubbing directions.
Notably, the Chinese to English dubbing direction appears to benefit more from joint modeling than the opposite direction, leading to a greater reduction in the MSE of the mel spectrogram.
This finding may also be related to the difference in the density of speaking style between the Chinese and English dubs
and highlights the potential effectiveness of joint training in both directions, particularly when the density of speaking styles in the source speech is lower than that in the target speech.

\section{Discussion}
\label{discussion}

Automatic dubbing is a potential research area in TTS and can be utilized to simplify the dubbing process
in real-world game and film production.
While several efforts have been made toward automatic dubbing and transferring the duration of each prosodic phrase to the dubbed language, no previous work has fully considered and transferred the speaking styles
in automatic dubbing.
In this paper, for the first time, a joint multiscale cross-lingual speaking style transfer framework is proposed
to simultaneously model the bidirectional speaking style transfer between two languages at both the global scale and local scale.
Extensive experiments show that the proposed framework is able to
jointly consider the speaking style of each word and the speaking style of the whole utterance in the source language
and to synthesize speech with appropriate speaking styles in the target language.

The difficulty of collecting speeches in different languages that are parallel in both meaning and speaking styles may explain why there is no approach to cross-lingual speaking style transfer modeling in automatic dubbing.
To address this challenge, we collected parallel speeches from the dubs in the video games.
These speeches are performed by experienced voice actors and recorded with professional equipment, making them ideal for learning speaking style transfer patterns across different languages.
However, as these speeches can only be employed in private, there may be a barrier to reproducing our work.
We urge game companies to publicly release their dubs in different languages to support future research on automatic dubbing.

To demonstrate the need for multiscale cross-lingual speaking style transfer in automatic dubbing, we conducted subjective observations on the cross-lingual speaking style transfers in our collected gaming corpus. The observations revealed a strong correlation between the speaking styles of parallel utterances in different languages. Specifically, at the utterance level, the global speaking styles of these parallel utterances are always the same or similar. At the word level, the local speaking styles of each word can often be \textit{matched} to a word in the other language, which means that these words have both the same or similar semantic information and local speaking styles.

Based on our observations, we propose a novel framework to jointly model cross-lingual speaking style transfer on both the global scale and local scale.
The proposed framework predicts the global speaking style for the dubbed language using global multimodal features from the source language and global textual features from the dubbed language.
At the local scale, we employ a bidirectional attention mechanism to transfer multimodal information from the source language to the dubbed language based on the attention weights learned between the semantic meanings of words in both languages.
Additionally, we use multitask learning to unify the learning of cross-lingual speaking style transfer for both dubbing directions and jointly optimize the attention weights at the local scale.
Although the proposed framework is initially designed and optimized on our gaming corpus,
it could also be generally employed for other dubbing scenarios, such as audiobook dubbing, as long as the dubbing processes still 
adhere to the observed phenomena on the gaming corpus.

However, the effectiveness of the proposed framework may be limited by the size of the corpus.
The corpus collected for our research is quite small, and we believe that our design could yield a better performance with a larger corpus.
The proposed framework may not work well when there are additional requirements during the dubbing process, such as matching the facial movements of the characters in the corresponding video.

\section{Conclusion}
\label{conclusion}

In this paper,
to properly transfer the speaking styles at both
the global scale (i.e., utterance level) and local scale (i.e., word level)
in automatic dubbing,
we propose a joint multiscale cross-lingual speaking style transfer framework
to model the speaking style transfer between languages at both scales with multitask learning.
The global and local speaking styles in each language
are extracted and utilized to predict the corresponding global and local speaking styles in the other language,
which are further synthesized to speech by MST-FastSpeech 2.
The effectiveness of our proposed framework is demonstrated by experiments and ablation studies.

\bibliography{references}
\bibliographystyle{IEEEtran}

\vfill

\end{document}

%% file: table/observation.tex
\begin{table}[tb]
  \caption{Aggregated responses in the subjective observation.}
  \label{tab:observation1}
  \centering
  \begin{tabular}{lccc}
    \toprule
    \textbf{Question} & \textbf{English} & \textbf{Chinese}\\
    \midrule
    \textbf{Has a \textit{special} global speaking style?} & $76.1\%$ & $78.1\%$\\
    \textbf{Are the global speaking styles similar?} & \multicolumn{2}{c}{$96.8\%$}\\
    \textbf{Has any \textit{special} local speaking style?} & $83.1\%$ & $83.7\%$\\
    \textbf{Percentage of \textit{matched} words?} & $86.7\%$ & $87.1\%$\\
    \bottomrule
  \end{tabular}
\end{table}

\begin{table}[tb]
  \caption{Pearson correlations of various audio properties between parallel utterances and matched words in two languages.}
  \label{tab:observation2}
  \centering
  \begin{tabular}{lccc}
    \toprule
    \textbf{Property} & \textbf{Utterance level} & \textbf{Word level}\\
    \midrule
    \textbf{Speaking rate} & $0.782$ & $0.482$\\
    \textbf{Pitch mean} & $0.850$ & $0.623$\\
    \textbf{Pitch std.} & $0.837$ & $0.245$\\
    \textbf{Energy mean} & $0.621$ & $0.414$\\
    \textbf{Energy std.} & $0.691$ & $0.261$\\
    \bottomrule
  \end{tabular}
\end{table}

%% file: table/results2.tex
\begin{table*}[tb]
  \caption{Mean squared errors of mel spectrograms at full band and the highest and lowest 10 dimensions.}
  \label{tab:results2}
  \centering
  \begin{tabular}{lcccc}
    \toprule
    \textbf{Speaking style transfer method} & \textbf{Direction} &\textbf{MSE (Mel)} & \textbf{MSE (High)} &\textbf{MSE (Low)} \\
    \midrule
    \textbf{No speaking style transfer \cite{ren2020fastspeech}} &\multirow{3}{*}{en-zh}& $4.694$ & $3.756$& $4.230$ \\
    \textbf{Duration transfer \cite{9747158}}&& $3.695$ & $3.079$& $3.133$ \\
    \textbf{Multi-scale speaking style transfer (Proposed)}&& $\mathbf{1.392}$& $\mathbf{1.090}$& $\mathbf{1.293}$ \\
    \midrule
    \textbf{No speaking style transfer \cite{ren2020fastspeech}} &\multirow{3}{*}{zh-en}& $5.385$& $4.305$& $4.627$ \\
    \textbf{Duration transfer \cite{9747158}}&& $3.301$& $2.697$& $2.593$ \\
    \textbf{Multi-scale speaking style transfer (Proposed)}&& $\mathbf{2.140}$& $\mathbf{1.623}$& $\mathbf{2.080}$ \\
    \bottomrule
  \end{tabular}
\end{table*}

\begin{table*}[tb]
  \caption{Subjective evaluations for different approaches.}
  \label{tab:results}
  \centering
  \begin{tabular}{lccccc}
    \toprule
    \textbf{Speaking style transfer method} & \textbf{Direction} & \textbf{MOS (Naturalness)} &\textbf{MOS (Transfer)} & \textbf{Preference rate}\\
    \midrule
    \textbf{No speaking style transfer \cite{ren2020fastspeech}} &\multirow{3}{*}{en-zh} & $2.761\pm0.097$& $3.161\pm0.075$ & $7.75\%$\\
    \textbf{Duration transfer \cite{9747158}}&& $3.562\pm0.099$& $3.923\pm0.073$ & $26.10\%$\\ 
    \textbf{Multi-scale speaking style transfer (Proposed)}&& $\mathbf{3.788\pm0.088}$ & $\mathbf{4.123\pm0.071}$ & $\mathbf{66.15\%}$\\
    \midrule
    \textbf{No speaking style transfer \cite{ren2020fastspeech}} &\multirow{3}{*}{zh-en}& $3.473\pm0.090$ & $3.358\pm0.077$ & $14.95\%$\\
    \textbf{Duration transfer \cite{9747158}}&& $3.615\pm0.093$ & $3.992\pm0.069$ & $40.72\%$\\
    \textbf{Multi-scale speaking style transfer (Proposed)}&& $3.626\pm0.097$& $4.005\pm0.073$ & $\mathbf{44.33\%}$\\
    \bottomrule
  \end{tabular}
\end{table*}

\begin{table*}[tb]
  \caption{Ablation studies on multiscale speaking style transfer and multitask learning on both directions.}
  \label{tab:ablation}
  \centering
  \begin{tabular}{lcccccc}
    \toprule
    \textbf{Approach} & \textbf{Direction} &\textbf{MSE (GST)} & \textbf{MSE (LST)} &\textbf{MSE (Mel)} & \textbf{MSE (High)} &\textbf{MSE (Low)} \\
    \midrule
    \textbf{Proposed} &\multirow{5}{*}{en-zh}&$4.220\times 10^{-3}$& $7.983\times 10^{-3}$ & $\mathbf{1.392}$ & $\mathbf{1.090}$& $1.293$ \\
    \textbf{-without local scale}& & $3.274 \times 10^{-3}$& -& $1.675$ & $1.545$& $1.333$ \\
    \textbf{-without global scale}&& - & $7.373\times 10^{-3}$& $1.660$ & $1.291$& $1.536$ \\
    \textbf{-without zh-en direction}&& $4.228\times 10^{-3}$& $8.097\times 10^{-3}$& $1.402$& $1.135$& $\mathbf{1.270}$ \\
    \midrule
    \textbf{Proposed} &\multirow{5}{*}{zh-en}& $2.012\times 10^{-3}$& $1.075\times 10^{-2}$& $\mathbf{2.140}$ & $\mathbf{1.623}$& $\mathbf{2.080}$ \\
    \textbf{-without local scale}&& $1.401\times 10^{-2}$& - & $2.699$ & $2.678$& $2.184$ \\
    \textbf{-without global scale}&& - & $1.092\times 10^{-2}$& $2.666$ & $2.053$& $2.533$ \\
    \textbf{-without en-zh direction}&& $2.024\times 10^{-3}$& $1.080\times 10^{-2}$& $2.368$& $1.804$& $2.272$ \\
    \bottomrule
  \end{tabular}
\end{table*}